\documentclass[aps,prl,twocolumn,floatfix,showpacs,superscriptaddress,nofootinbib]{revtex4-2}
\usepackage{amsmath,amsfonts,amssymb,bm}
\usepackage{graphicx}
\usepackage{color}
\usepackage{subfigure}
\usepackage{multirow}
\usepackage{textcomp}
\usepackage{slashed}
\usepackage{bbm}
\usepackage[bottom]{footmisc}
\usepackage{parskip}
\definecolor{purple}{rgb}{0.5,0,0.5}
\definecolor{blue}{rgb}{0.0,0,1.0}
\usepackage[colorlinks=true, pdfstartview=FitV, linkcolor=purple, citecolor= purple, urlcolor=blue]{hyperref}

\newcommand{\al}{\alpha}

\newcommand{\beq}{\begin{equation}}
\newcommand{\eeq}{\end{equation}}
\newcommand{\ba}{\begin{array}}
\newcommand{\ea}{\end{array}}
\newcommand{\bea}{\begin{align}}
\newcommand{\eea}{\end{align}}
\newcommand{\bi}{\begin{itemize}}
\newcommand{\ei}{\end{itemize}}
\newcommand{\ben}{\begin{enumerate}}
\newcommand{\een}{\end{enumerate}}
\newcommand{\bc}{\begin{center}}
\newcommand{\ec}{\end{center}}
\newcommand{\bl}{\begin{flushleft}}
\newcommand{\el}{\end{flushleft}}
\newcommand{\br}{\begin{flushright}}
\newcommand{\er}{\end{flushright}}

\newcommand\Eqn[1]{Eq.~(\ref{#1})}  

\renewcommand{\>}{\rangle}   

\newcommand\comment[1]{ \hbox{[{\it Comment suppressed here.}\/]} }
\newcommand\hide[1]{}
\newcommand{\skipover}[1]{}

\begin{document}
\title{The Dichotomous Nature of the $\sigma$ Meson and the Nucleon D-Term}

\author{Zanbin Xing}
\affiliation{School of Physics, Nankai University, Tianjin 300071, China}

\author{Kh\'epani Raya}
\affiliation{Department of Integrated Sciences and Center for Advanced Studies in Physics, Mathematics and Computation, University of Huelva, E-21071 Huelva, Spain.}

\author{Yu-xin Liu}
\affiliation{Department of Physics, Peking University, Beijing 100871, China}

\author{Lei Chang}%
\affiliation{School of Physics, Nankai University, Tianjin 300071, China}

\date{\today}

\begin{abstract}
Employing a symmetry-preserving contact-interaction formulation of the Dyson-Schwinger equations in quantum chromodynamics (QCD), we examine the identity of the $\sigma$ meson and its implications for the gravitational structure of hadrons. In this framework, the scalar meson emerges as the chiral partner of the pion, with both states' properties tightly connected to the mechanisms of mass generation in QCD. We find that, above a critical coupling that triggers dynamical chiral symmetry breaking, the D-terms of the constituent quark, the pion, and $\sigma$ saturate at fixed values $D_{q,\pi,\sigma}=-1/3,-1,-7/3$. By examining the coupling strength evolution of the D-terms, this pattern follows naturally once a dual nature for the $\sigma$ meson is recognized: it behaves both as a quark–antiquark composite and as a dilaton arising from spontaneous scale symmetry breaking. This unified picture yields the prediction $D_N\sim -3$ for the nucleon D-term, consistent with lattice QCD, dispersive analyses, and diverse continuum studies.
\end{abstract}

\maketitle

\noindent
\noindent\emph{1.\ Introduction}\,---\,%
A central challenge in contemporary hadron physics is to understand how mass is generated within the strong-interactions sector of the Standard Model, quantum chromodynamics (QCD). \emph{Strong} mass generation accounts for nearly $99\%$ of the mass of visible matter, most of which resides in protons and neutrons. The proton stands out as the most fundamental composite state produced by QCD, and, arguably, its mass sets a characteristic dynamical scale of the theory\,\cite{BMW:2008jgk,Roberts:2020hiw,Achenbach:2025kfx}. Nevertheless, a comprehensive picture begins with the analysis of the lightest hadron in Nature: the pion. Its significance trascends beyond the role as an effective mediator of the nuclear strong force. The pion can be understood as conventional quark–antiquark bound state and as the (pseudo) Nambu–Goldstone (NG) boson associated with dynamical chiral symmetry breaking (CSB)\,\cite{Klevansky:1992qe,Nambu:1961tp,Goldstone:1962es}. This dual nature makes the pion indispensable in providing a valuable understanding of the origin of mass in the SM, e.g.\,\cite{Horn:2016rip,Roberts:2021nhw,Raya:2024ejx}. The picture is expanded by also considering the scalar meson $\sigma:=f_0(500)$. Such a state remains a long-standing enigma in strong interaction physics\,\cite{Caprini:2005zr,Pelaez:2015qba,Rupp:2018azz}, one whose resolution may offer further insights into the emergence of mass\,\cite{Scadron:1982eg,Delbourgo:1982tv,Hatsuda:1985ey,Kunihiro:1995rb}.

In conventional quark models that account for dynamical CSB, such as the Nambu–Jona–Lasinio framework,\,\cite{Nambu:1961tp,Klevansky:1992qe}, the scalar–isoscalar channel arises naturally as the chiral partner of the pion, interpreted as a correlated exchange of pairs in the scattering process by Nambu. Nevertheless, in the real world, its exceptionally large width,~\cite{ParticleDataGroup:2024cfk,Caprini:2005zr}, and its prominent role in $\pi\pi$ scattering, e.g.\,\cite{Garcia-Martin:2011iqs,Cao:2023ntr,Pelaez:2024uav}, challenge a simple quark–antiquark interpretation. Furthermore, like the pion, the $\sigma$ is unusually light, with $m_\sigma \sim400-500$ MeV. These features raise the question of whether this state might be a (pseudo) NG boson associated with a broken symmetry,\,\cite{Isham:1970gz,Ellis:1971sa,Crewther:1970gx,Crewther:1971bt,Bardeen:1985sm}. The answer may lie in chiral effective theories{\color{red},\,\cite{Crewther:2015dpa}}, where the $\sigma$ emerges as the lightest scalar–isoscalar excitation and naturally connects to the trace anomaly and the gravitational form factors (GFF), e.g.\,\cite{Zwicky:2023fay,Stegeman:2025sca}. The latter are actual probes of the energy-momentum tensor and thus scale symmetry\,\cite{Kobzarev:1962wt,Pagels:1966zza,Lorce:2025oot,Polyakov:2018zvc}. Closely related near-conformal scenarios similarly predict a light dilatonic scalar whose mass is controlled by the smallness of the QCD $\beta$-function in the infrared \cite{Golterman:2016lsd,Zwicky:2023krx,Zwicky:2023bzk}. Moreover, recent lattice QCD explorations of the pion and nucleon GFFs,\,\cite{Hackett:2023nkr,Hackett:2023rif}, are found compatible with scalar meson dominance,\,\cite{Broniowski:2024oyk,Broniowski:2025ctl,Stegeman:2025sca}, and a dilatonic view of the $\sigma$ meson\,\cite{Stegeman:2025sca}. Altogether, these developments support a description of the $\sigma$ as the (pseudo)–NG boson associated with spontaneous scale symmetry breaking (SSB).

In light of the above, we investigate the gravitational properties of the pion and the $\sigma$ meson, specifically the so-called D-term, and we also take a look at the nucleon case. The nucleon D-term is considered the \emph{last unknown global property}, yet across hadrons it may be characterized as the \emph{least known global property}. Herein we discuss hadron D-terms from two perspectives: one from bound state viewpoint which is realized through a symmetry-preserving Dyson-Schwinger equation (DSE) framework characterized by a vector-vector contact interaction (CI)~\cite{Gutierrez-Guerrero:2010waf,Xing:2022jtt}. This framework captures the dichotomous nature of the pion, both as a quark–antiquark bound state and as the NG mode of dynamical CSB. The other standpoint is dilaton theory\,\cite{Zwicky:2023fay}. Notably, a striking level of consistency between the two perspectives is achived once the $\sigma$ meson is identified as dilaton. Such consistency further allows us to provide a simple but promising estimation for the nucleon D-term.

\noindent\emph{2.\ Setting the stage}\,---\,%
The hadron GFFs follow from the matrix elements of the QCD energy–momentum tensor, e.g.\,\cite{Polyakov:2018zvc}. For a spinless meson of mass $M_0$, the relevant matrix element reads:
\begin{equation}
    \<P_+|T_{\mu\nu}|P_-\>=\frac{\delta^T_{\mu\nu}}{2}D(Q^2)+P_\mu P_\nu A(Q^2)+2M^2_0\delta_{\mu\nu}\bar{c}(Q^2)
\end{equation}
where $\delta^T_{\mu\nu}=Q_\mu Q_\nu-\delta_{\mu\nu}Q^2$, and $P_\pm=P\pm Q/2$ denote the outgoing (+) and incoming (-) meson momenta. The scalar functions $A(Q^2),D(Q^2),\bar{c}(Q^2)$, define the corresponding GFFs. This work focuses on the D-term, which is defined by $D:=D(Q^2=0)$.
To investigate pion and $\sigma$ meson GFFs, with particular emphasis on the D-term, we build upon the framework developed in Ref.~\cite{Xing:2022mvk,Sultan:2024hep}. All components are determined explicitly using the CI model, which incorporates the leading-order symmetry-preserving kernels, the so-called rainbow-ladder truncation\,\cite{Munczek:1994zz,Bender:1996bb}, and a momentum-independent gluon propagator. Accordingly, the interaction kernel governing the quark, meson and the meson-meson scattering amplitude assumes the form:
\begin{equation}
K=\frac{4\pi\alpha_{\text{eff}}}{m_{G}^{2}}\gamma_{\alpha}\otimes\gamma_{\alpha}\,;
\end{equation}
here \(m_{G} = 0.5\ \text{GeV}\) plays the role of an effective gluon mass, and \(\alpha_{\text{eff}}\) denotes the dimensionless coupling strength. This choice is motivated by the observed saturation of the QCD coupling in the infrared region~\cite{Deur:2025idg}, where the interaction becomes effectively momentum-independent, signaling a conformal-like region at low-energies. Such a minimalistic framework preserves essential non-perturbative facets of QCD, while also offering significant calculational advantages. In particular, by focusing on the zero momentum-transfer, it provides an ideal setting for evaluating form factors\,\cite{Raya:2021pyr,Hernandez-Pinto:2023yin,Hernandez-Pinto:2024kwg}.

A key element to compute the GFFs is the proper identification of the meson-meson scattering amplitude, which can be splitted naturally into contributions from intrinsic quark degrees of freedom and those associated with intermediate bound states, most notably the scalar channel~\cite{Xing:2022mvk,Sultan:2024hep}. This decomposition is particularly valuable for understanding the D-term, as the scalar channel is essential for reproducing the characteristic chiral-limit value \(D_\pi = -1\), expected from the soft-pion theorem\,\cite{Polyakov:1999gs,Novikov:1980fa,Voloshin:1982eb}. Furthermore, despite being important in the description of electromagnetic form factors, vector meson poles are found to contribute nothing directly to the GFFs\,\cite{Xing:2022mvk,Sultan:2024hep,Yao:2024ixu,Xu:2023izo}. This observation reinforces the notion that scalar dynamics play a central role in constraining the D-term of hadrons.

By identifying the meson–meson scattering amplitude, we avoid the conventional route of evaluating hadron GFFs via the dressed quark–graviton vertex (QGV)\,\cite{Yao:2024ixu,Xu:2023izo}. Even so, the dressed-quark GFFs remain accessible through the dressed QGV, which in the CI framework is given by:
\begin{equation}
\label{eq:QGV}
    \Gamma^G_{\mu\nu}(k_+,k_-)=\gamma^G_{\mu\nu}(k_+,k_-)+\left[\frac{\delta^T_{\mu\nu}}{4M_q} D_q(Q^2)+\delta_{\mu\nu}\bar{c}_q(Q^2)\right]
\end{equation}
where $\gamma^G_{\mu\nu}(k_+,k_-)$ is identifiable form QCD's EMT and guarantees the fulfillment of the relevant Ward-Green-Takahashi identities\,\cite{Brout:1966oea}. This same structure appears as the inhomogeneous term in the BSE for $\Gamma^G_{\mu\nu}$, which can be solved in close analogy with the treatment of the meson–meson scattering amplitude,\cite{Xing:2022mvk,Sultan:2024hep}. Also in analogy, $\gamma^{G}_{\mu\nu}(k{+},k_{-})$ carries the contribution from intrinsic quark degrees of freedom, while the expression in square brackets reveals the presence of the scalar bound–state pole.

\begin{figure}[ht]
    \centering
    \includegraphics[width=\linewidth]{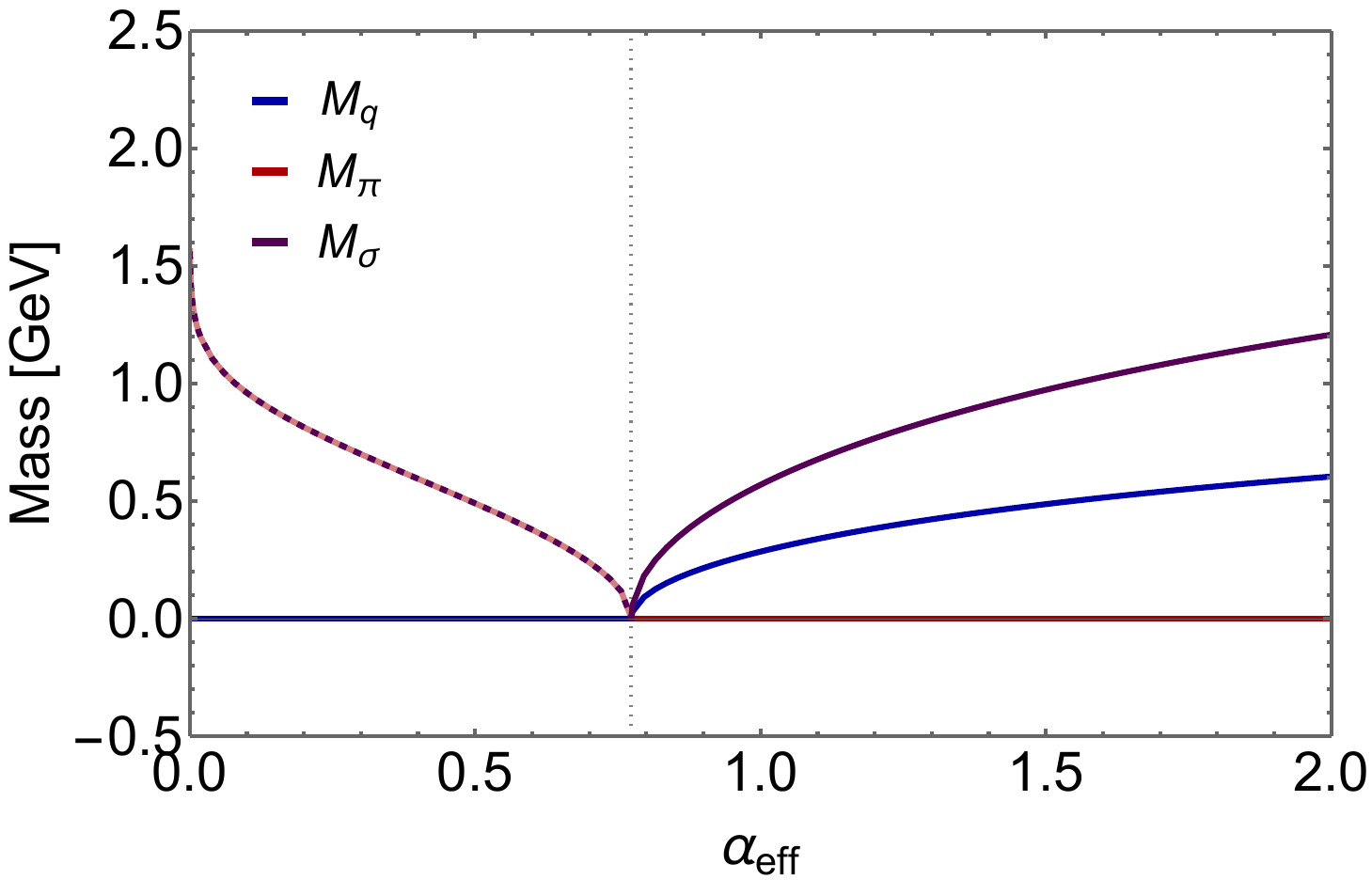}
    \caption{Dependence of the dressed-quark, pion, and $\sigma$ meson masses on the effective coupling $\alpha_{eff}$. The vertical dotted line marks the critical coupling $\alpha_c$. Below this threshold, pion and $\sigma$ are degenerate states.}
    \label{fig:mass}
\end{figure}

\noindent\emph{3.\ Phase transition and $\sigma$'s dichotomy}\,---\,%
It is well-established that QCD's strong interactions possess a critical coupling strength \(\alpha_c\) that governs the onset of dynamical CSB~\cite{Hatsuda:1985ey,Vogl:1989ea,Roberts:1994dr,Wang:2012me,Raya:2013ina,Williams:2007ey}. When $\alpha_{\text{eff}}$ exceeds a critical value $\alpha_{c}$, the system transitions from the symmetric Wigner mode to the symmetry-broken NG phase, dynamically generating a characteristic quark mass scale $M_q$ (see Fig.~\ref{fig:mass}). To be specific, let us express the DSE of the quark propagator in the chiral limit as follows:
\begin{equation}\label{eq:gap}
    [1-\mathcal{K}]M_{q}=0\,.
\end{equation}
Here $M_{q}$ is the quark mass function, which in principle depends on the quark momentum, and $\mathcal{K}$ is the one-body kernel, generally expressed in terms of the external and internal momenta, as well as the mass function itself. Although the mass function is momentum-independent in the CI model , and so the operator $[1-\mathcal{K}]$ is merely a multiplicative factor, the form in the Eq.\,\eqref{eq:gap} remains fully general and does not affect our subsequent discussion on criticality. That said, the solution of the quark DSE has three possible configurations depending the coupling strength $\alpha_{\text{eff}}$:
\begin{subequations}
    \begin{align}
    \alpha_{\text{eff}}>\al_c:&\,1-\mathcal{K}=0\, \& \,M_{q}\neq0\,,\\
    \alpha_{\text{eff}}<\al_c:&\,1-\mathcal{K}\neq0 \,\& \,M_{q}=0\,,\\
    \alpha_{\text{eff}}=\al_c:&\,1-\mathcal{K}=0 \,\& \,M_{q}=0\,.
\end{align}
\end{subequations}

Above the critical strength, the physical solution of \Eqn{eq:gap} is the positive Nambu mode $M_q>0$, determined by the dynamical part of the gap equation, $1-\mathcal{K}=0$. This condition not only generates a dynamical quark mass, thereby breaking chiral symmetry, but also ensures the emergence of a massless pseudoscalar bound state in the associated BSE\,\cite{Maris:1997hd}. In other words, it triggers the appearance of the pion as the NG boson of dynamical CSB\,\,\cite{Munczek:1994zz,Bender:1996bb}.

Below the critical strength, the condition $1-\mathcal{K}\neq0$ holds for any real $M_q$, so the quark DSE admits only the trivial solution $M_q=0$. Chiral symmetry is therefore not dynamically broken and the two Nambu solutions vanish rather than merging smoothly into the Wigner mode, e.g. Refs.\,\cite{Wang:2012me,Raya:2013ina,Williams:2007ey}. In addition, while $1-\mathcal{K}\neq0$ prevents a massless solution for the corresponding BSE, a massive pseudoscalar bound state would still exist, and would be degenerate with its chiral partner. This pattern is evident from Fig.\,\ref{fig:mass}. The physical interpretation of such a pseudoscalar is unclear, but it cannot be associated with a NG boson: in the absence of dynamical CSB, no massless NG mode can emerge.

Finally, interesting phenomena occur at the exact critical coupling $\alpha_{\text{eff}}=\alpha_c$. First of all, the two Nambu modes collapse to the Wigner mode because the solution of the  $1-\mathcal{K}=0$ condition is just $M_q=0$. Chiral symmetry ($M_q=0$), and the condition that drives its dynamical breaking ($1-\mathcal{K}=0$), are thus realized simultaneously, allowing the formation of a massless pseudoscalar bound state. If this state is identified as the NG boson of dynamical CSB, the theory appears to retain and break chiral symmetry at once. This situation is reminiscent of a coexistence point, much like water and ice exist simultaneously at the critical temperature. Moreover, at the critical point, the $\sigma$ meson and the pion become mass-degenerate. This behavior has been previously observed,~\cite{Chang:2008ec}, suggesting a more general pattern rather than an artifact of the present model.

\begin{figure}[ht]
    \centering
    \includegraphics[width=\linewidth]{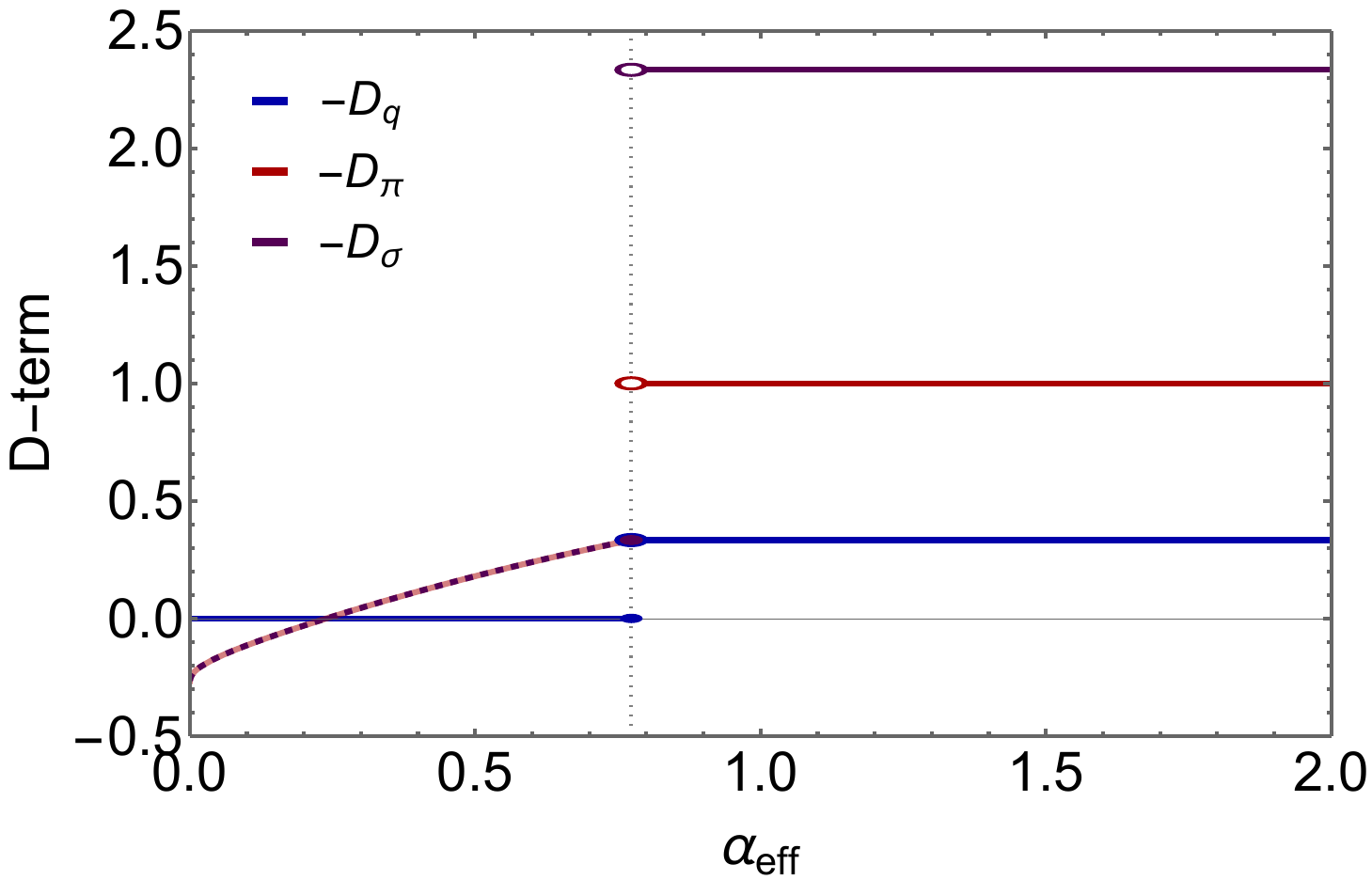}
    \caption{Dependence of the dressed-quark, pion, and $\sigma$ meson D-terms on the effective coupling $\alpha_{eff}$. At the critical point, corresponding to the vertical dotted line, one finds: $D_q=0$, and $D_\pi=D_\sigma=-1/3$. Below $\alpha_c$, pion and $\sigma$ become degenerate states.}
    \label{fig:Dterm}
\end{figure}

The very existence of the pion as a (pseudo) NG boson requires working in the regime where the effective coupling is strong enough to trigger dynamical CSB. Once this condition is met, our framework automatically reproduces the chiral-limit condition $D_\pi=-1$ (see Fig.\,\ref{fig:Dterm}), thereby confirming both its symmetry-preserving construction and its ability to capture the essential infrared dynamics of QCD. Remarkably, the same setup that encodes the dual nature of the pion also generates an accompanying scalar quark–antiquark bound state\,\cite{Roberts:2011cf}. Although this state cannot account for the broad scalar resonance observed experimentally, it can nonetheless be identified with the $\sigma$ meson, endowing it with a distinct physical interpretation. To delve into the properties of the scalar mode and, in particular, its connection to the pion, as well the associated symmetries and breaking patterns, we now analyze how key quantities evolve as a function of the effective coupling, including the dressed quark, pion and $\sigma$ meson masses, and corresponding D-terms.

Our investigation reveals a striking universal behavior in the gravitational response of these composite systems. As illustrated in Fig.~\ref{fig:Dterm}, once the coupling exceeds the critical value and so the system enters the Nambu phase, the D-terms of all three objects saturate to constant values: the constituent quark settles at \(D_q = -1/3\), the pion retains its characteristic \(D_{\pi} = -1\), and the $\sigma$ meson acquires the larger value $D_\sigma=-7/3$. The fact that these results remain unchanged as the coupling varies suggests that they may represent fundamental properties of these entities in the symmetry-broken phase. While the pion's D-term invariance at \(-1\) is naturally understood through the soft-pion theorem, a reflection of its NG boson character and independent of interaction details in the chiral limit, the parallel invariance exhibited by the dressed quark (\(-1/3\)) and \(\sigma\) meson (\(-7/3\)) leads us to our central question: why do these quantities \emph{freeze out} into seemingly universal values while other structural properties, like masses, decay constants, in-hadron condensates, evolve significantly with the coupling? 

Furthermore, as revealed by Fig.~\ref{fig:Dterm}, the D-term behavior exhibits a visible discontinuity exactly at the critical coupling $\alpha_c$. When this value is approached as \(\alpha_{\text{eff}} \to \alpha_c^-\), the dressed-quark, whose mass is driven to zero at $\alpha_c$, retains a vanishing D-term. In contrast, both the pion and the $\sigma$ meson approach the common value $-1/3$, despite also being massless and degenerate at the transition point. This behavior naturally leads to a further conceptual question: in the vicinity of chiral restoration, where these objects are massless yet the interaction strength remains finite, how should we understand the emergence of different D-term values for fermionic and bosonic states? Such a situation situation resembles a soft version of fermion–boson duality\,\cite{Chen:2017lkr}: their masses align at criticality, yet the internal structure encoded in the D-term keeps track of their distinct spin and structure.

A possible resolution to these interrogants arise once the $\sigma$ meson’s dual role is recognized: both as a $q\bar{q}$ bound state and as a (pseudo) NG boson of spontaneous SSB. As addressed below, this would be in accordance with the $\sigma$ meson phase transition, as well as scale symmetry and its breaking, thereby motivating a dilaton-based picture.

We now remark that above the critical point, beyond ensuring dynamical CSB and therefore the emergence of a massless pseudoscalar meson, the present framework also produces:
\begin{equation}
m_{\sigma} = 2M_q\,.
\label{eq:mass_relation}
\end{equation}
This chiral limit mass relation is another invariant relation that persists in the symmetry-broken Nambu phase~\cite{Nambu:1961tp,Hatsuda:1985ey,Klevansky:1992qe,Roberts:2011cf,Scadron:1982eg,Delbourgo:1982tv}. Analogously to the pion amplitude in the Goldberger-Treiman relation $\Gamma_{\pi}(P^{2}=0)\sim i\gamma_{5}M_{q}$,\,\cite{Maris:1997hd}, the scalar amplitude exhibits the behavior $\Gamma_{\sigma}(P^{2}=-4M_{q}^{2})\sim M_{q}$. 
This parallelism indicates that both the pion and its chiral partner are strongly tied by the dynamical piece of the gap equation, $1-\mathcal{K}=0$.
The relation in Eq.\,\eqref{eq:mass_relation} reveals another important aspect of scale symmetry in this framework: above the critical strength, where the quark acquires a non-zero constituent mass, scale symmetry is \emph{explicitly} broken by the dynamical CSB-triggered $M_q$. This entails that the $\sigma$ meson is massless only if chiral symmetry is not broken, either dynamically or explicitly. Consequently, in the Nambu phase, the $\sigma$ meson is generated by dynamical SSB (embodied in $1-\mathcal{K}=0$), along with an explicit source ($M_q\neq0$). In this sense, it is most naturally interpreted as a pseudo NG boson of spontaneous SSB. Moreover, \Eqn{eq:mass_relation} corresponds to a situation of zero binding energy: the mass of the bound state is exactly the sum of the masses of its two dressed-quark constituents, with no negative binding energy in the usual sense. A relation this rigid (completely independent of the coupling strength) is unlike what one expects from ordinary dynamical binding mechanisms. Instead, it resembles the behavior of a state sitting at a symmetry-imposed critical point, or the manifestation of a deeper underlying identity in the theory. At a greater extent, this behavior cannot be easily accommodated with a purely bound state interpretation.

As the coupling reaches its critical value, the system reaches a transition where the correlation length diverges, signalling the emergence of scale invariance. Since no dynamical mass is generated at this point, there is no term that would explicitly violate scale symmetry. In close analogy with chiral symmetry, the onset of dynamical SSB becomes visible exactly at criticality through the appearance of a massless scalar bound state. Such a system would play the role of a dilaton, the NG mode associated with spontaneous SSB, providing a consistent interpretation of the scalar channel within the CI model. 

Below the critical coupling, the scale symmetry remains intact because of the massless constituent quark. However, instead of massless NG bosons due to the spontaneous breaking of scale and chiral symmetry, a pair of massive degenerate bound states emerge in this phase, resulting from the failure of satisfying the dynamical equation $1-\mathcal{K}=0$. The physics in this phase needs further investigation and is beyond the scope of this work.

With these observations in mind, we now consider the D-term from the perspective of dilaton theory. For clarity, we adopt an alternative but equivalent expressions to those derived in Ref.~\cite{Zwicky:2023fay}, at leading-order within dilaton chiral perturbation theory. The D-terms of the pion, the dilaton, the generic scalar state, and the spin-$1/2$ fermion thus read:
\begin{subequations}
    \begin{align}
    \label{eq:dtermsa}
    D_{\pi}(q^2)&=-\frac{1}{d-1}+\frac{2(d_\varphi m_D^2+(\Delta_{q\bar{q}}-2d_\varphi )m_\pi^2)}{(d-1)(q^2-m_D^2)}\,,\\
    \label{eq:dtermsb}
    D_{D}(q^2)&=-\frac{1}{d-1}+\frac{2(c_3-3d_\varphi)m_D^2}{(d-1)(q^2-m_D^2)}\,,\\
    \label{eq:dtermsc}
    D_{\varphi}(q^2)&=-\frac{1}{d-1}+\frac{2(2m_\varphi^2+d_\varphi m_D^2)}{(d-1)(q^2-m_D^2)}\,,\\
    \label{eq:dtermsd}
    D_{\psi}(q^2)&=0+\frac{4m_\psi^2}{(d-1)(q^2-m_D^2)}\,.
    \end{align}
\end{subequations}
Here $d$ is the spacetime dimension, $d_\varphi=(d-2)/2$ is the dimension of a free scalar field $\varphi$, and $c_3$ denotes coefficient of the cubic term in the dilaton potential. The quantity $\Delta_{q\bar{q}}$ represents the scaling dimension of the operator $q\bar{q}$, determined as $\Delta_{q\bar{q}}=d-2$ to ensure the soft pion theorem,\,\cite{Zwicky:2023fay,Zwicky:2023krx,Zwicky:2023bzk}. Expressed in this way, the $-1/(d-1)$ and $0$ terms on the the right hand side coincide precisely with the intrinsic quark contribution to the meson and quark D-terms at the critical coupling, while the remaining components arise from the dilaton pole contributions. One may even speculate that, in analogy with the quark case, that the nucleon D-term likewise receives no intrinsic contribution and is instead saturated by pole terms alone. This would align with the view that the fermion D-term is of purely dynamical origin\,\cite{Polyakov:2018zvc,Lorce:2025oot,Hudson:2017oul}.

By identifying the $\sigma$ meson as a dilaton, the consistency between \Eqn{eq:dtermsb} and ~\Eqn{eq:dtermsc} requires $c_3 = 2(d-1)$. Considering then a four-dimensional spacetime, and employing the chiral limit relation from Eq.\,\eqref{eq:mass_relation}, the D-terms for the massless and massive dilaton are found to be:
\begin{equation}
\left(D_{q},D_{\pi},D_{\sigma}\right)=\left(0,-\frac{1}{3},-\frac{1}{3}\right)_\text{(massless dilaton)}\,,
\end{equation}
and
\begin{equation}
\left(D_q,D_{\pi},D_{\sigma}\right)=\left(-\frac{1}{3},-1,-\frac{7}{3}\right)_\text{(massive dilaton)}\,.
\end{equation}
These results are in perfect alignment with the expectations of the CI model. A key technical point in the massless regime is the sequence of limits: one must first impose $m_D = m_\varphi = 0$ and only then take the limit $q^2\to0$. With this prescription in place, once the $\sigma$ is interpreted as the dilaton and the improved energy–momentum tensor is consistently incorporated, the formerly enigmatic \emph{frozen} D-term values acquire a transparent interpretation. In this manner, the present framework not only reproduces the universal constants found in the Nambu phase, but also naturally accounts for the discontinuity at the critical coupling $\alpha_{c}$.

\noindent\emph{4.\ Nucleon's D-term}\,---\,%
Building on \Eqn{eq:dtermsd}, which is applicable to all spin-$1/2$ fermions, we can now address the nucleon D-term. As noted earlier in this manuscript, the bulk of the nucleon mass is generated dynamically, as a reflection of the emergent facets of QCD. Given that this mass is approximately the sum of the constituent-quark masses, it is natural to adopt $M_{N}\sim 3 M_q$ as a first approximation. This simple observation leads to the following result:
\begin{equation}
    D_{N}\sim -\frac{4}{3}\frac{(3M_{q})^{2}}{(2M_{q})^{2}} = -3\,.
\end{equation}
Notably, this seemingly naive estimate aligns with contemporary results from lattice QCD~\cite{Hackett:2023rif}, continuum approaches~\cite{Yao:2024ixu}, and dispersive analyses~\cite{Cao:2025dkv}. In particular, lattice QCD yields $D_{N} = -3.87(97)$ and $-3.35(58)$ from dipole and z-expansion fits, respectively;  continuum studies report $D_{N} = -3.114(10)$ and dispersive methods find $D_N = -3.38_{-0.35}^{+0.34}$. Notably, the scalar meson dominance yields $D=-3.0(4)$\,\cite{Broniowski:2025ctl}, and the dilaton picture examined thoroughly in Ref.\,\cite{Stegeman:2025sca} supports a $\sigma$-meson contribution to the nucleon's D-term of  $D_N^\sigma=-3.01(39)$.

The remarkable consistency between our direct evaluation and sophisticated assessments highlight the reliability of our approach. This agreement is not accidental as it follows from a framework that faithfully encodes essential aspects of symmetries and their breaking patterns, together with the identification of the $\sigma$ meson as a dilaton. As a result, even with its apparent simplicity, the present framework reproduces the correct infrared behavior of QCD, with remaining deviations attributable only to subleading interaction details.

\noindent\emph{5.\ Conclusions}\,---\,%
In this work we investigated the gravitational structure of the pion, $\sigma$ meson, and the nucleon, capitalizicing on their D-terms. Our analysis roots on the CI model, which features a rainbow-ladder truncation of QCD's DSEs, and a zero-range effective interaction. Despite its relative simplicity, this framework captures crucial non-perturbative characteristics of QCD, including the appearance of a massless pion as the NG boson of dynamical CSB, and the chiral limit prescription $D_\pi=-1$. This same approach naturally places the $\sigma$ meson as the chiral partner of the pion, and closely defines its mass as the sum of those of its constituent quarks. Consequently, the mass of the scalar meson would be set at a great extent by the strength of dynamical CSB.

A notable feature of our analysis is that once the effective coupling surpasses the critical value $\alpha_c$ and dynamical CSB sets in, the D-terms of the quark, the pion, and the $\sigma$ meson lock onto universal constants: $D_q=-1/3$, $D_\pi=-1$, $D_\sigma=-7/3$. Below this threshold, the pion and the $\sigma$ meson become degenerate states, and their D-terms acquire a dependence on the coupling strength. Interestingly, exactly at the critical point, chiral symmetry behaves like both preserved and broken: it is retained with vanishing quark mass, however, massless pseudoscalar and scalar states are produced, and these share the common D-term value $-1/3$ at $\alpha_c^{-}$, while the D-term splits at $\alpha_c^{+}$. This observation suggests that the scalar meson may emerge as a NG boson associated with the breaking of some symmetry. Adopting a dilaton viewpoint supports this intuition and a transparent explanation for the fixed D-term values identified earlier. In this sense, and in close analogy to the pion, the $\sigma$ meson adopts a dual character, and may be regarded simultaneously as a quark-antiquark bound state and the (pseudo) NG boson of spontaneous SSB. Thus, at the exact critical coupling, the pion and $\sigma$ meson are massless and behave as true NG boson modes of their associated symmetries. Above this point, the strength of dynamical CSB sets the mass of the scalar meson, and so the explicit source of SSB.

By keeping the dual interpretation of the scalar meson, and noting that the nucleon mass is approximately the sum of its constituent-quark masses, we arrive at the prediction $D_N=-3$. Remarkably, this seemingly rudimentary estimate aligns with state-of-the-art calculations from different backgrounds. Thus, the dichotomous character of the $\sigma$ meson arises as a natural bridge between spontaneous SSB and the gravitational structure of hadrons, in particular the part encoded within the D-term. Ultimately, the validity of this picture must be settled empirically. High-precision determinations of the nucleon D-term at the forthcoming Electron-Ion Collider,~\cite{Anderle:2021wcy,Accardi:2012qut} will play a valuable role.

\section{Acknowledgments} We thank Fei Gao and Roman Zwicky for valuable constructive comments on the manuscript. Work supported by: National Natural Science Foundation of China, grant no. 12135007; Postdoctoral Fellowship Program of CPSF under Grant Number GZC20240759; Spanish Ministry of Science and Innovation (MICINN grant no.\ PID2022-140440NB-C22); and Junta de Andalucía (grant no.\ P18-FR-5057). 

\bibliography{main}

\end{document}